\documentclass[twocolumn,showpacs]{revtex4}

\usepackage{graphicx}%
\usepackage{dcolumn}%
\usepackage{amsmath}%

\begin{document}
%\DeclareGraphicsExtensions{.jpg,.pdf,.mps,.png}

\date{18 June 2001}

\title{Crystal Structure and Magnetism of the Linear-Chain Copper Oxides 
Sr$_5$Pb$_{3-x}$Bi$_{x}$CuO$_{12}$}

\author{K. Yamaura}
\email[E-mail:]{YAMAURA.Kazunari@nims.go.jp}
\affiliation{Advanced Materials Laboratory, National Institute for Materials
Science, 1-1 Namiki, Tsukuba, Ibaraki 305-0044, Japan}
\affiliation{Japan Science and Technology Corporation, Kawaguchi, Saitama 
332-0012, Japan}

\author{Q. Huang}
\affiliation{NIST Center for Neutron Research, National Institute of Standards
and Technology, Gaithersburg, Maryland 20899}
\affiliation{Department of Materials and Nuclear Engineering, University of 
Maryland, College Park, Maryland 20742}

\author{E. Takayama-Muromachi}
\affiliation{Advanced Materials Laboratory, National Institute for Materials
Science, 1-1 Namiki, Tsukuba, Ibaraki 305-0044, Japan}

\begin{abstract}
The title quasi-1D copper oxides ($0 \leq x \leq 0.4$) were investigated
by neutron diffraction and magnetic susceptibility studies.
Polyhedral CuO$_4$ units in the compounds were found to comprise linear-chains
at inter-chain distance of approximately 10 \AA.~~
The parent chain compound ($x =$ 0), however, shows less anisotropic magnetic
behavior above 2 K, although it is of substantially antiferromagnetic ($\mu_
{\rm eff}$ = 1.85 $\mu_{\rm B}$ and $\Theta_{\rm W} =$ -46.4 K) spin-chain 
system.
A magnetic cusp gradually appears at about 100 K in $T$ vs $\chi$ with the Bi
substitution.
The cusp ($x =$ 0.4) is fairly characterized by and therefore suggests the 
spin gap nature at $\Delta/k_{\rm B} \sim 80$ K.
The chain compounds hold electrically insulating in the composition range.
\end{abstract}

\pacs{75.50.-y, 81.40.Rs}

%75.50.-y Studies of specific magnetic materials 
%75.50.Ee Antiferromagnetics
%75.50.Mm Magnetic liquids
%81.40.Rs Electrical and magnetic properties

\maketitle

\pagebreak

\section{Introduction}

A variety of Cu(II) based linear-chain oxides is known such as Ca$_4$Cu$_5$O$_
{10}$, Li$_2$CuO$_2$, CuGeO$_3$, Sr$_2$CuO$_3$, SrCuO$_2$,
and Y$_2$Ca$_2$Cu$_5$O$_{10}$, and indeed one dimensional
Heisenberg antiferromagnetism was observed in those materials \cite{RMP98MI}.
Studies on the copper oxides play an important 
role in research on the one dimensional electronic systems, raising
and settling fundamental issues as to the nature of quasi-particles and 
electron correlations in condensed matter \cite{RMP98MI}.
We have recently been studying copper based chain materials in order to find
additional systems showing correlations among their magnetic and electrical 
properties and crystal structure.

In this paper we report the crystal structure and magnetic properties of the 
linear-chain copper oxides Sr$_5$Pb$_{3-x}$Bi$_x$CuO$_{12}$ ($0\leq x\leq0.4$)\@.
In the parent compound Sr$_5$Pb$_3$CuO$_{12}$, distorted CuO$_4$ units are 
linearly connected by sharing a conner-oxygen \cite{JSSC90JSK,JSSC91TGNB}.
One-dimensionally anisotropic magnetic properties were then expected to 
Sr$_5$Pb$_3$CuO$_{12}$ because the copper ion has 
spin-1/2 electronic configuration and the principal magnetic bond 
Cu--O--Cu ($\sim$ 165$^\circ$) in chain is supposed antiferromagnetic \cite
{PRB98YM}.
Experimental studies, however, seemed not to be focused on magnetism of this
compound probably because insufficient quality of the so-far samples and relatively high 
degree of structural disorder \cite{JSSC90JSK,JSSC91TGNB}.
Even the most high quality sample was made at copper-rich starting composition
Sr:Pb:Cu=3:2:1 rather than copper-stoichiometric composition \cite{JSSC91TGNB}.
The sample quality level was acceptable for a regular structural study, but rather
unsuitable for detailed magnetic studies because the sample should have unknown magnetic
impurities \cite{JSSC91TGNB}.
The magnetism of the quasi-1D compound Sr$_5$Pb$_3$CuO$_{12}$ remained ambiguous.

In this study, further improvement of quality of the copper-stoichiometric sample
of Sr$_5$Pb$_3$CuO$_{12}$ and partial Bi substitution for Pb, i.e. Sr$_5$Pb$_{3-x}
$Bi$_x$CuO$_{12}$, up to $x=$ 0.4 were achieved in single-phase polycrystalline form, 
followed by investigations by neutron 
diffraction and magnetic susceptibility studies.
We found the quasi-1D compound Sr$_5$Pb$_3$CuO$_{12}$ shows Curie-Weiss-type 
magnetism even at low temperature ($>$ 2 K), and at the composition Sr$_5$Pb$_
{2.6}$Bi$_{0.4}$CuO$_{12}$ a finite energy gap ($\Delta$/k$_{\rm B} \sim$ 80
K) in magnetic system was suggested.

\section{Experimental}

The polycrystalline single-phase samples were prepared by high-temperature 
ceramic synthesis technique.
Mixtures of SrCO$_3$, PbO, Bi$_2$O$_3$, and CuO with the ratio Sr:Pb:Bi:Cu
= 5:$3-x$:$x$:1 ($x =$ 0, 0.1, 0.2, 0.3, 0.4, 0.6, 0.8, 1.0) were heated 
at 800 $^\circ$C for 15 hours in air.
After the initial treatment, grinding and heating in air were repeated at 800
$^\circ$C\@, followed by those at 850 and 900 $^\circ$C for 95 hours in total.
The samples were then ground and molded into pellets, and again heated in air
at 950 $^\circ$C for 45 hours.
The heating at 950 $^\circ$C in either nitrogen or oxygen instead of air 
results in poor quality of final productions.
Subsequently, the sintered pellets were annealed at 500 $^\circ$C in 
compressed (100 MPa) gas, 20 \% oxygen in argon, for 5 hours.
The quality of the final productions were quite sensitive to the highest heating 
temperature.
Dense alumina crucibles were employed to hold the samples in the synthesis procedure.
The magnetic properties of the samples were studied by a commercial apparatus
between 2 and 390 K\@.
The highest applied magnetic field was 55 kOe.
All of the pellets thus obtained were electrically highly insulating.

\begin{figure*} 
\includegraphics{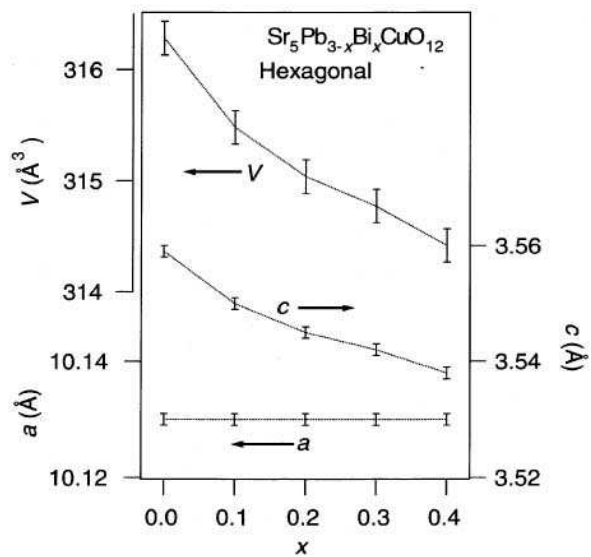}
\caption{Lattice constants and unit cell volume of Sr$_5$Pb$_{3-x}$Bi$_x$CuO$_
{12}$~(0 $\leq x \leq$ 0.4, {\it P}\={6}2{\it m}) measured by means of powder x-ray
diffraction at room temperature.}
\label{Lattice}
\end{figure*}

Powder x-ray diffraction with CuK$\alpha$ radiation at room temperature was
employed to characterize the samples and found a part of those ($x=0-0.4$) was single-phase
within the sensitivity of the technique ($\sim$ 1 \%)\@.
Composition dependence of lattice constants and volume of the hexagonal unit
cell are plotted in Fig.\ref{Lattice}.
Linear dependence in $V$, $c$ vs $x$, and $x$ independence of $a$ are clearly 
seen, indicates the solid solution is formed.
The end members, Sr$_5$Pb$_3$CuO$_{12}$ and Sr$_5$Pb$_{2.6}$Bi$_{0.4}
$CuO$_{12}$, were then selected for further crystal structure studies by 
neutron diffraction at 10 and 295 K\@.
The neutron diffraction data were obtained by the BT-1 high-resolution powder
diffractometer at the NIST Center for Neutron Research.
A Cu(311) monochromator was employed to produce a coherent neutron beam
($\lambda = 1.5401$ \AA) with 15$^{\prime}$, 20$^ {\prime}$, and 7$^{\prime}
$ collimators before and after the monochromator, and after the sample, 
respectively.
The intensity of diffracted neutron beam was measured between 3 and 160 
degrees at each 0.05 degrees step in 2$\Theta$.
Crystal structure parameters of the copper oxides were then refined to a high
degree of agreement by calculations with the intensity and angle data on the
program GSAS \cite{LANLR90ACL}\@.
Neutron scattering amplitudes for the elements in the refinements were 
set to 0.702, 0.940, 0.853, 0.772, and 0.581 ($\times$10$^{-12}$) cm for Sr,
Pb, Bi, Cu, and O\@, respectively \cite{LANLR90ACL}\@.

\section{Results and Discussions}
\subsection{Crystal Structure}
\begin{figure*}
\includegraphics{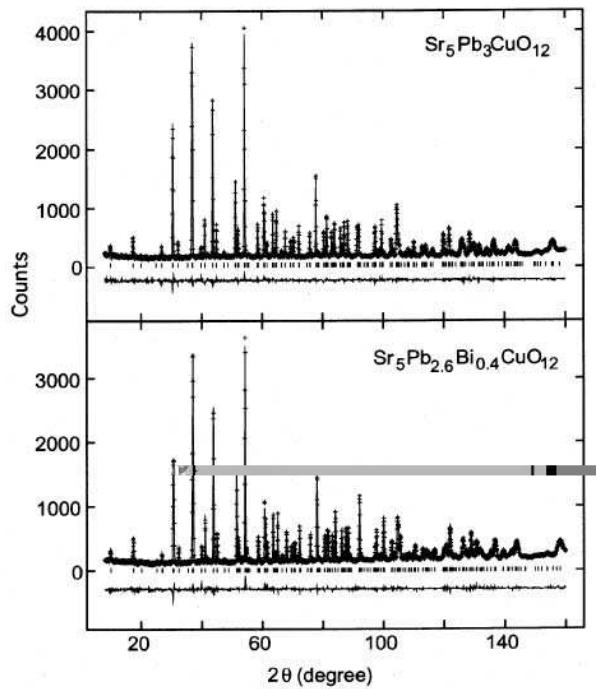}
\caption{Observed (crosses) and calculated (solid curve) powder neutron 
diffraction profiles (295 K) of Sr$_5$Pb$_3$CuO$_{12}$ and Sr$_5$Pb$_{2.6} 
$Bi$_{0.4}$CuO$_{12}$.~~The small vertical bars indicate calculated positions
for the nuclear Bragg reflections. The lower part of each panel shows 
difference between the profiles.}
\label{NDP}
\end{figure*}

At first, oxygen stoichiometry and crystal structure of the $x =$ 0 sample were
investigated.
A crystal structure model [{\it P}\={6}2{\it m}, $a =$ 10.1089(6) \AA~and $c
=$ 3.5585(2) \AA] previously proposed for Sr$_5$Pb$_3$CuO$_{12}$, of which the
powder sample was prepared at the composition Sr:Pb:Cu=3:2:1 \cite{JSSC91TGNB}, was tested.
In the top panel in Fig.\ref{NDP}, the observed profile is presented with the
calculated one.
As you can see, all observed peaks are clearly reproduced at 4--5 \% levels 
in agreement factors, indicating the qualities of the sample and the 
refinement.
The structural model is valid for the present compound prepared at the correct composition.  
Lattice constants of the hexagonal unit cell ({\it P}\={6}2{\it m}) are $a =$
10.1297(2) \AA~and $c =$ 3.559 80(7) \AA~for Sr$_5$Pb$_3$CuO$_{12}$ at 295 
K\@;
there are insignificant difference between lattice parameters for the 321 and 
present samples (0.2 \% and 0.04 \% in $a$ and $c$ parameters, respectively).
The structure parameters are summarized in Table \ref {table}.
The occupancy factors of oxygen in the normally occupied sites, O(2) and O(3), 
were refined to within one standard deviation of 1.00, and the sites were 
therefore taken to be fully occupied.
The occupancy factors in the partially occupied sites, O(1), O(4) and O(5), 
are slightly larger than 2/3, 1/12 and 1/12, respectively, which are expected
when the compound is oxygen-stoichiometric at 12 moles of oxygen per formula unit.
Magnetometric study of the sample indicates the valence state of copper is 
fairly close to +2 as discussed later, supports the oxygen-stoichiometric 
composition Sr$_5$Pb$_3$CuO$_{12}$. 
Because the occupancy factors are highly correlated with the thermal-displacement
parameters, which are unusual due to the remarkably low levels
of oxygen occupancy, and complicated local structure in chain, the oxygen 
occupancy factors are presumed slightly overestimated.
Unreasonable values for $B_{11}$ of O(4) were probably gained due to the same reasons.
The composition of the sample is then Sr$_5$Pb$_3$CuO$_{12}$ or possibly somewhat 
oxygen-superstoichiometric.

The neutron diffraction profile of the $x = 0.4$ sample at 295 K is presented
in the bottom panel in Fig.\ref{NDP} as well as that of $x = 0$ sample.
The obtained structure parameters at 295 and 10 K are listed and compared with those for 
the $x=$ 0 sample in Table \ref{table}.
The qualities of the sample of Sr$_5$Pb$_{2.6}$Bi$_{0.4}$CuO$_{12}$ and the
refinement are as good as those achieved for Sr$_5$Pb$_3$CuO$_{12}$.
Lattice constants of the hexagonal unit cell ({\it P}\={6}2{\it m}) are $a =$
10.1236(2) \AA~and $c =$ 3.541 82(6) \AA~for Sr$_5$Pb$_{2.6}$Bi$_{0.4}$CuO$_
{12}$ at 295 K, and $a =$ 10.1042(2) \AA~and $c =$ 3.534 90(6) \AA~at 10 
K\@.
The oxygen quantity is independent on the Bi doping level; no significant 
deference in oxygen occupancy factors was detected between both the samples.
On the assumption that Pb and Bi valences are +4 and +3, respectively, the 
formal copper valence of Sr$_5$Pb$_{2.6}$Bi$_{0.4}$CuO$_{12}$ should be +2.4, does 
not meet the expectation from the observed effective Bohr magneton as shown later. 
Partial increment of the Bi valence to +5 therefore probably occur.

\begin{figure*} 
\includegraphics{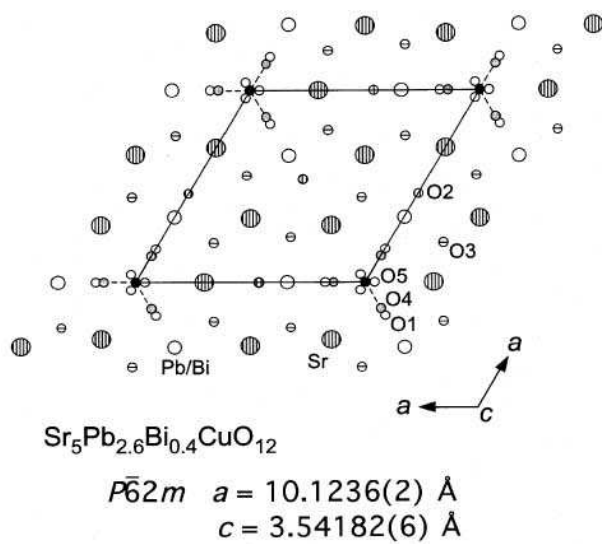}
\caption{Schematic crystal structure view for Sr$_5$Pb$_{2.6}$Bi$_{0.4}$CuO$_
{12}$ drawn from the structure parameters at 295 K. The hexagonal unit cell 
is indicated by the solid lines. Copper ions are located at the corners of the
unit cell in this view.}
\label{Structure}
\end{figure*}

\begin{figure*} 
\includegraphics{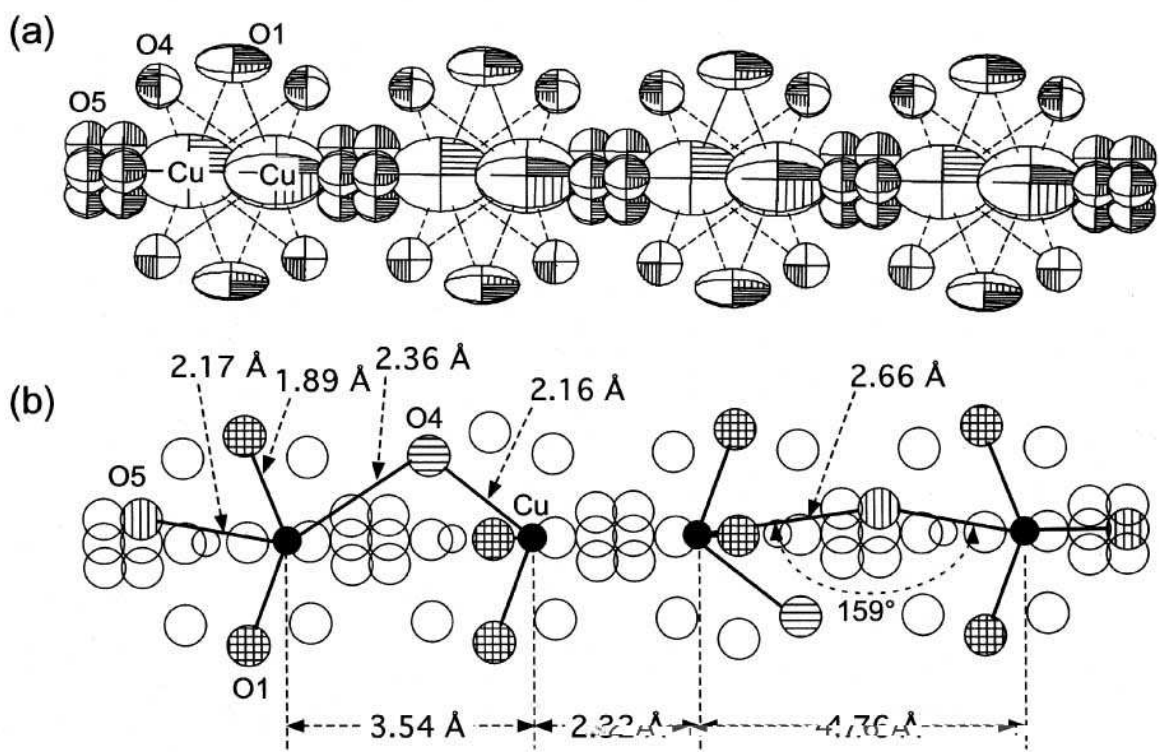}
\caption{(a) Schematic thermal ellipsoids view for the copper-oxygen chain 
along {\it c}--axis, drawn from the structure parameters of Sr$_5$Pb$_{2.6} 
$Bi$_{0.4}$CuO$_{12}$ at 295 K, showing the randomly occupied oxygen positions
O(1), O(4) and O(5) around copper, (b) and a probable local coordination between 
copper and oxygen within the average structural model.}
\label{Chain}
\end{figure*}

Schematic structural view for Sr$_5$Pb$_{2.6}$Bi$_{0.4}$CuO$_{12}$ is 
presented as a representative of the solid solution in Fig.\ref{Structure}, based
on the structure parameters at 295 K.
The Bi doped compound retains the linear-chain structure basis, in which 
copper--oxygen polyhedra form chains at inter-chain distance of approximately
10 \AA\@.
The local structure of the chain is highly complicated as indicated in Fig.\ref{Chain}a, 
where all possible oxygen and copper positions are indicated.
In Fig.\ref{Chain}b, probable local arrangements of CuO$_4$ units include a 
misordering of copper atom (third one from the left side) is shown by precluding
coordinations of copper and oxygen resulting in intolerably short bond
distances.
On the other hand, the 4--18 and 2--10 \AA$^2$ thermal displacement parameters
for Cu and O(1), respectively, are also unusual, 
indicating probable presence of local displacements of Cu and O(1) 
along $c-$axis (Fig.\ref{Chain}a) and frequent occurrence of the irregular
copper atom arrangement in local.
Due to the complicated chain structure, a comparison between the structure 
data at 10 and 295 K for Sr$_5$Pb$_{2.6}$Bi$_{0.4}$CuO$_{12}$ does not give 
a significant contribution to detect possible local lattice distortions 
associated with magnetism as found in CuGeO$_3$ \cite{PRL94JPP}.
Further crystal structure investigations into Sr$_5$Pb$_{2.6}$Bi$_{0.4}$CuO$_
{12}$ include microscopic studies and structure modulation analysis should
be interesting.

\subsection{Magnetic Properties}

\begin{figure*} 
\includegraphics{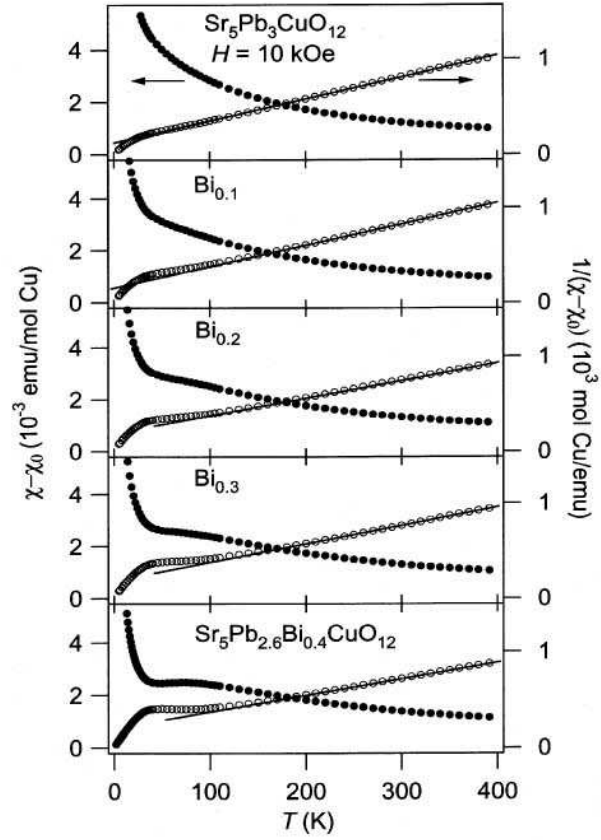}
\caption{Temperature dependence of magnetic susceptibility of Sr$_5$Pb$_{3-x}
$Bi$_x$CuO$_{12}$~(0 $\leq x \leq$ 0.4), and inverse plots of the 
susceptibility. The data were obtained at 10 kOe on heating after cooling each
sample without the applied magnetic field. The solid lines indicate fits to 
the Curie-Weiss law.}
\label{Magnetic}
\end{figure*}

\begin{figure*} 
\includegraphics{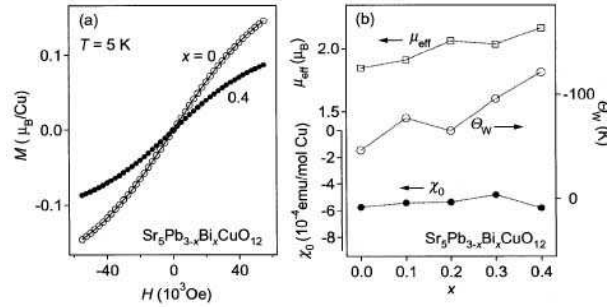}
\caption{Applied magnetic field dependence of magnetization of 
Sr$_5$Pb$_3$CuO$_{12}$ and Sr$_5$Pb$_{2.6}$Bi$_{0.4}$CuO$_{12}$ at 5 K.~~
(b) Magnetic parameters of Sr$_5$Pb$_{3-x}$Bi$_x$CuO$_{12}$ (0 $\leq x \leq$
0.4) estimated by fitting the Curie-Weiss law to the magnetic 
susceptibility data.}
\label{CW}
\end{figure*}

The temperature dependence of the magnetic susceptibility of Sr$_5$Pb$_{3-x}
$Bi$_{x}$CuO$_{12}$ ($0 \leq x \leq 0.4$) was measured at magnetic field of 
10 kOe.
Those data are presented in Fig.\ref{Magnetic} as $\chi-\chi_0$ vs. $T$ and
$1/(\chi-\chi_0)$ vs. $T$ plots ($\chi_0$ is temperature-independent portion). 
The magnetic field dependence of the magnetization at 5 K of the selected samples
($x=$0 and 0.4) are shown in Fig.\ref{CW}a.
In the enough high temperature region above 200 K, the magnetic data are 
expected free from any of magnetic ordering contributions either long or short
range because the characteristic magnetic cusp was found far below the
temperature region.
The Weiss temperature ($\Theta_{\rm W}$), $\chi_0$, and the effective Bohr 
magneton ($\mu_{\rm eff}$) were estimated by fitting the Curie-Weiss law to
the data above 200 K\@.
The solid lines in Fig.\ref{Magnetic} indicate the fits, and those magnetic
parameters estimated are presented in Fig.\ref{CW}b.
The formula for the fitting by least-squares method was
\begin{eqnarray}
\chi(T)={N\mu^2_{\rm eff} \over 3k_{\rm B}(T-\Theta_{\rm W})}+\chi_{\rm 0}~~
(T > 200~{\rm K}),
\label{CW200}
\end{eqnarray}
where $k_{\rm B}$ and $N$ are Boltzmann and Avogadro's constants, 
respectively.

The Weiss temperatures estimated for Sr$_5$Pb$_{3-x}$Bi$_x$CuO$_{12}$
are characteristic of antiferromagnetic interactions (Fig.\ref{CW}b),
and the absolute value tends to be gradually elevated as the Bi 
doping level increases.
The $\chi_0$ in the composition range is almost constant, $\sim-6\times10^{-4}$ 
emu/mol Cu, which is remarkable
approximately one magnitude larger than the
values for the other chain structure copper oxides \cite{RMP98MI}.
This fact is probably due to the unusually large number of diamagnetic 
elements per copper (20 moles per mole Cu) \cite{OUP32JHVV}, therefore we 
decided not to make further concerns on this issue. 
The effective Bohr magneton is slightly above 1.73 $\mu_{\rm B}$, calculated
from the ideal electronic configuration of Cu(II) (3$d^9$, $t_{\rm 2g}^6$ $e_{\rm g}^3$, 
$S = 1/2$) using the formula $\mu_{\rm eff}=2\sqrt{S(S+1)}~~\mu_{\rm B}$.
The effective $g$-value are expected different somewhat rather than 2.00 if
the valence of copper is fixed at +2.00 in the composition range. 
By considering the negative value of Weiss temperature and the effective Bohr
magneton, it is concluded that antiferromagnetic interaction is dominant 
between the nearest neighbor spin--1/2 magnetic moments in chain of 
Sr$_5$Pb$_{3-x}$Bi$_{x}$CuO$_{12}$.
Ferromagnetic interactions in Sr$_5$Pb$_{3-x}$Bi$_{x}$CuO$_{12}$ are 
insignificant because ferromagnetic characters such as hysteresis and 
spontaneous moments were not found at all in the $M$ vs. $H$ curves (Fig.\ref{CW}a).

\begin{figure*} 
\includegraphics{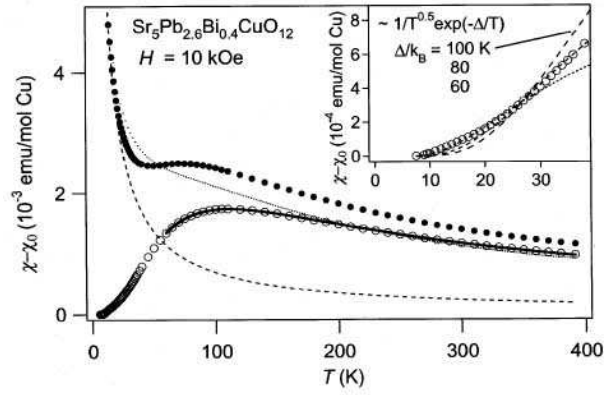}
\caption{Quantitative analysis of the magnetic susceptibility data of 
Sr$_5$Pb$_{2.6}$Bi$_{0.4}$CuO$_{12}$. The open circles result from subtraction
of the low-temperature Curie component (broken curve, 13.3 \% level) from 
original data (closed circles). Fat solid curve indicates a fit to 
alternating linear-chain model using a least squares method at {\it J} 
$_1$/k$_{\rm B} $= 183 K, $\alpha$ = 0.58, and {\it g} = 2.22. Inset: 
Estimation of magnitude of the possible magnetic gap by applying spin gap 
model \cite{PRB94MT}, indicating it is approximately 80 K. Temporally applied
an independent model, in which a linear relationship between Bonner-Fiser and
Curie components exists, is indicated by dotted curve in the main panel.}
\label{Bi04}
\end{figure*}

Magnetic susceptibility of the antiferromagnetic linear-chain system is divided
into three terms \cite{PRB00DCJ}:
\begin{eqnarray}
\chi(T)= (1-f)\chi_{\rm 1D}(T) + f\chi_{\rm C}(T) + \chi_0,
\label{chichichi}
\end{eqnarray}
where $f$ is level of the Curie term [$\chi_{\rm C}(T)$], and $\chi_{\rm 1D}
(T)$ stands for the intrinsic linear-chain magnetic susceptibility.
To further analyze the $x =$ 0.4 data, which are most influenced by the linear-chain
magnetism as the cusp appears at $\sim$ 80 K, the low temperature quasi-Curie
component was subtracted (Fig.\ref{Bi04}).
The Eq.\ref{CW200} with $\Theta_{\rm W} ^\prime$ instead of $\Theta_{\rm W}$ 
was employed to estimate the quasi-Curie component by fitting to the original data 
below 8 K as indicated by the broken curve in Fig.\ref{Bi04}
($f \sim$ 13.3 \% and $\Theta_{\rm W}^\prime = -0.66$ K\@).
As the chemical impurity level of the sample is less than 1 \%, it is 
concluded that the nearly 13 \% of magnetic moments in Sr$_5$Pb$_{2.6}$Bi$_
{0.4}$CuO$_{12}$ formally contribute to form the low temperature quasi-Curie term.
Origins to produce the almost free-spins, which yield the nearly zero Weiss 
temperature, are possibly associated with the local structural disorders, 
which may cut and cause end of chains.
The detailed analysis of the substantial amount of free-spins in chain using further 
experimental studies is left for future works.

After the Curie term subtraction, character of the $\chi_{\rm 1D}(T)$ term becomes clear;
it goes to zero on cooling.
At present we presume that the particular magnetism is due to an energy gap 
between magnetic ground and excited states, as found 
in the spin-ladder and the alternating-exchange linear-chain materials \cite
{RMP98MI,PRB00DCJ}.
The formula proposed for spin gap system at $T \ll \Delta/k_{\rm B}$ \cite
{PRB94MT}, 
\begin{eqnarray}
\chi_{\rm 1D}(T) \sim {1 \over \sqrt{T}}\exp{(-{\Delta \over T})},
\label{SG}
\end{eqnarray}
was applied on the data below 35 K, as shown in the inset in Fig.\ref {Bi04},
and found that the possible energy gap ($\Delta/k_{\rm B}$) is approximately 80 K.
An alternative fit to a model in which there are Curie-Weiss and Heisenberg 
Bonner-Fisher components \cite{PR64JCB} was unlikely at all. 
Even at the best result (dotted curve at $S=$1/2, $g=$ 2, $\Theta_{\rm W}^\prime =$ -0.19 K, 
$f=$ 23.6 \%, and $J_{\rm eff}/k_{\rm B}$= 72.4 K) it was too poor to reproduce 
the original data.    

Alternating antiferromagnetic spin-chain system has been intensively
investigated in (VO)$_2$P$_2$O$_7$ and Cu(NO$_3$)$_2\cdot$2.5H$_2$O \cite
{PRB00DCJ,PRB83JVCB,PRB87DCJ,PRB94TB}.
The model was tested on the present linear-chain material Sr$_5$Pb$_{2.6}$Bi$_
{0.4}$CuO$_{12}$ by applying the applicable formula to the data above 60 K, 
which has three independent variable parameters $J_1$, $\alpha$ ($=J_2/J_1$)
and $g-$value, where $J_1$ and $J_2$ are antiferromagnetic exchange constants
alternating along chain ($J_1 \ge J_2 \ge 0$) \cite {IC81JWH,JAP81WEH}.
The best fit was obtained at $J_1/k_{\rm B}=$ 183 K, $\alpha =$ 0.58 and $g 
=$ 2.22, as shown in Fig.\ref{Bi04} (fat solid curve).
Thus, the magnetic cusp temperature was calculated 115 K from the formula 
$\sim 0.63J_1/k_{\rm B}$, which is expected almost independent on $\alpha$ \cite 
{JAP79JCB}, and the spin gap 77 K from $\sim J_1-J_2$ \cite{PRB98PRH}, those
in fact match well with the observations.
The alternating spin chain model at $\alpha \sim 0.6$ is, therefore, the most
probable model to account for the character of $\chi_{\rm 1D}(T)$ of 
Sr$_5$Pb$_{2.6}$Bi$_{0.4}$CuO$_{12}$ if the spin gap nature is prime essential
in the magnetism of this compound.
Further investigations employing probes more sensitive to the microscopic 
magnetic environment, such as NMR studies, would play a significant role to figure out
the magnetic ground state and the excitation.
As a clue to further elucidate the possible spin gap nature of Sr$_5$Pb$_{2.6}
$Bi$_{0.4}$CuO$_{12}$, inelastic-neutron-scattering study would also be
expected particularly if a single crystal becomes available.

\section{Conclusion}
It was found that the substantial Bi substitution for Pb in Sr$_5$Pb$_3$CuO$_
{12}$ plays a key role to profile the one dimensional antiferromagnetic 
feature, without a significant influence on the electrical conductivity.
At $x =$ 0.4, the alternating magnetic interactions at $\alpha \sim 0.6$ was
strongly suggested, which potentially produce spin-singlet ground state with 
approximately 80 K spin-gap. 
There are two major models to explain a cusp in $T$ vs $\chi$ in
antiferromagnetic linear-chain system as far as we know.
As already described the one does not match ultimately with the present 
magnetic data, and the other one reaches sufficiently convincible level in the
fitting study; these facts imply the spin gap nature in the $x=$ 0.4 chain 
compound.
Because conclusive evidence is not provided yet about the probable spin gap,
further investigations include theoretical point of view into the Bi doped 
compound would be of interest.
The alteration possibly depend on the Bi doping level, which modifies the
local magnetic environments.
It is not figure out yet how the doping improves the local magnetic environment
around copper and what produces the low-temperature Curie term.
Further Bi doping would probably be favorable to obtain credible information about the
issues and make clear nature of the magnetism of the title spin chain 
compounds.
As an electrical carrier doping to spin gap system would be of great interest,
further studies of a variety of chemical substitutions for Sr$_5$Pb$_3$CuO$_
{12}$ are in progress.

\acknowledgments
We are grateful to M. Isobe (AML/NIMS) for helpful discussions.
This research was supported in part by the Multi Core Project administrated 
by Ministry of Education, Culture, Sports, Science, and Technology of Japan.

\begin{table*}
\caption{Structure parameters of Sr$_5$Pb$_3$CuO$_{12}$ at 295 K (first line), 
Sr$_5$Pb$_{2.6}$Bi$_{0.4}$CuO$_{12}$ at 295 K (second line), and Sr$_5$Pb$_
{2.6}$Bi$_{0.4}$CuO$_{12}$ at 10 K (third line). Space group for those is {\it
P}\={6}2{\it m}. The lattice parameters are {\it a} = 10.1297(2) {\AA} and {\it
c} = 3.559 80(7) {\AA} for Sr$_5$Pb$_3$CuO$_{12}$, {\it a} = 10.1236(2) {\AA}
and {\it c} = 3.541 82 (6) {\AA} for Sr$_5$Pb$_{2.6}$Bi$_{0.4}$CuO$_{12}$ at 295 
K, and {\it a} = 10.1042(2) {\AA} and {\it c} = 3.534 90(6) {\AA} for Sr$_5$Pb$_
{2.6}$Bi$_{0.4}$CuO$_{12}$ at 10 K.}
\label{table}
\begin{tabular}{llllllll}
Atom &Site &$x$         &$y$         &$z$         &$n$      &             &\\
\hline
Sr(1)&$2d$ &1/3         &2/3         &1/2         &1        &             &\\
     &     &1/3         &2/3         &1/2         &1        &             &\\
     &     &1/3         &2/3         &1/2         &1        &             &\\
Sr(2)&$3g$ &0.7019(3)   &0           &1/2         &1        &             &\\
     &     &0.7017(3)   &0           &1/2         &1        &             &\\
     &     &0.7027(2)   &0           &1/2         &1        &             &\\
Pb/Bi&$3f$ &0.3407(2)   &0           &0           &1        &             &\\
     &     &0.3381(2)   &0           &0           &1        &             &\\
     &     &0.3378(2)   &0           &0           &1        &             &\\
Cu   &$2e$ &0           &0           &0.3360(39)  &0.5      &             &\\
     &     &0           &0           &0.3275(43)  &0.5      &             &\\
     &     &0           &0           &0.3351(37)  &0.5      &             &\\
O(1) &$3g$ &0.1747(6)   &0           &1/2         &0.771(17)&             &\\
     &     &0.1767(5)   &0           &1/2         &0.785(16)&             &\\
     &     &0.1777(4)   &0           &1/2         &0.755(14)&             &\\
O(2) &$3g$ &0.4619(3)   &0           &1/2         &1        &             &\\
     &     &0.4614(3)   &0           &1/2         &1        &             &\\
     &     &0.4610(2)   &0           &1/2         &1        &             &\\
O(3) &$6j$ &0.2371(2)   &0.4432(3)   &0           &1        &             &\\
     &     &0.2369(2)   &0.4435(3)   &0           &1        &             &\\
     &     &0.2374(1)   &0.4439(2)   &0           &1        &             &\\
O(4) &$6i$ &0.1383(13)  &0           &0.2471(41)  &0.109(4) &             &\\
     &     &0.1384(17)  &0           &0.2085(75)  &0.105(4) &             &\\
     &     &0.1424(13)  &0           &0.2396(48)  &0.107(3) &             &\\
O(5) &$6i$ &0.9592(11)  &0           &0.9238(35)  &0.109(4) &             &\\
     &     &0.9587(14)  &0           &0.9300(89)  &0.105(4) &             &\\
     &     &0.9588      &0           &0.9301      &0.107(3) &             &\\
\\
      &$B_{11}$ (\AA$^2$)&$B_{22}$ (\AA$^2$)&$B_{33}$ (\AA$^2$) 
      &$B_{12}$ (\AA$^2$)&$B_{13}$ (\AA$^2$)&$B_{23}$ (\AA$^2$)&\\
\hline
Sr(1) &1.43(10)  &$= B_{11}$[Sr(1)]&1.56(14)  &0.717(49)  &0 &0 &\\
      &1.43(11)  &$= B_{11}$[Sr(1)]&1.25(15)  &0.719(54)  &0 &0 &\\
      &0.861(78) &$= B_{11}$[Sr(1)]&0.54(11)  &0.430(39)  &0 &0 &\\
Sr(2) &1.30(9)   &0.723(97)   &0.825(99) &0.362(49)  &0 &0 &\\
      &1.16(8)   &0.83(10)    &0.85(10)  &0.414(51)       &0 &0 &\\
      &0.702(64) &0.427(84)   &0.506(83) &0.213(42)       &0 &0 &\\
Pb/Bi &0.487(61) &0.962(90)   &1.73(6)   &0.481(45)  &0 &0 &\\
      &0.629(57) &0.873(88)   &1.19(6)   &0.437(44)       &0 &0 &\\
      &0.386(46) &0.369(69)   &0.804(47) &0.184(34)       &0 &0 &\\
Cu    &9.39(61)  &$= B_{11}$(Cu)&13.4(14)&4.70(31)   &0 &0 &\\
      &9.24(65)  &$= B_{11}$(Cu)&17.9(20)&4.62(32)        &0 &0 &\\
      &8.36(50)  &$= B_{11}$(Cu)&16.1(16)&4.18(25)        &0 &0 &\\
O(1)  &3.66(31)  &3.09(38)    &8.11(56)  &1.54(19)   &0 &0 &\\
      &2.46(24)  &2.20(33)    &9.33(59)  &1.10(17)        &0 &0 &\\
      &2.41(21)  &1.66(28)    &6.57(42)  &0.83(14)        &0 &0 &\\
O(2)  &1.74(11)  &2.51(14)    &0.38(11)  &1.26(7)    &0 &0 &\\
      &1.43(11)  &2.99(15)    &0.63(12)  &1.50(8)         &0 &0 &\\
      &0.743(82) &1.69(11)    &0.473(98) &0.845(54)       &0 &0 &\\
O(3)  &0.838(87) &1.41(10)    &1.45(8)   &0.402(95)  &0 &0 &\\
      &0.820(84) &1.61(10)    &1.34(8)   &0.591(90)       &0 &0 &\\
      &0.362(67) &0.770(85)   &0.960(65) &0.033(69)       &0 &0 &\\
O(4)  &-0.91(42) &1.33(53)    &2.16(69)  &0.67(26)   &2.46(33) &0&\\
      &0.21(45)  &4.63(86)    &6.7(15)   &2.32(43)   &2.23(61) &0&\\
      &-0.36(35) &2.27(54)    &3.49(82)  &1.13(27)   &2.13(30) &0&\\
O(5)  &$=B_{11}$[O(4)]&$= B_{22}$[O(4)]&$= B_{33}$[O(4)]&$= B_{12}$[O(4)]&$= B_{13}$[O(4)]&$= B_{23}$[O(4)]&\\
      &$=B_{11}$[O(4)]&$= B_{22}$[O(4)]&$= B_{33}$[O(4)]&$= B_{12}$[O(4)]&$= B_{13}$[O(4)]&$= B_{23}$[O(4)]&\\
      &$=B_{11}$[O(4)]&$= B_{22}$[O(4)]&$= B_{33}$[O(4)]&$= B_{12}$[O(4)]&$= B_{13}$[O(4)]&$= B_{23}$[O(4)]&\\
\\
 &$R_{\rm p} =$&4.03 \%&$R_{\rm wp} =$&4.90 \%&$\chi^{2} =$&1.144&\\
 &             &4.35 \%&              &5.30 \%&            &1.184&\\
 &             &4.34 \%&              &5.23 \%&            &1.510&\\
\end{tabular}
\end{table*}

\end{document}